# Barrier efficiency of sponge-like $La_2Zr_2O_7$ buffer layers for YBCO-coated conductors

*Leopoldo Molina [1,*], Haiyan Tan [1], Ellen Biermans [1], Kees J. Batenburg [2], Jo Verbeeck [1], Sara Bals [1] and Gustaaf Van Tendeloo [1]*

[1] EMAT, University of Antwerp

Groenenborgerlaan 171,

BE-2020 Antwerp, Belgium

[2] Vision Lab, University of Antwerp

Universiteitsplein 1,

BE-2020 Wilrijk, Belgium

\* Corresponding author. E-mail adress: leopoldo.molina-luna@ua.ac.be

Tel: + 32 3 265 32861; Fax: + 32 3 265 3318

**Keywords**

 Coated conductors, thin films, HAADF-STEM, EELS, Tomography

**PACS**

 68.37.Ma, 68.37.Lp, 78.67.Rb, 61.05.Tv, 79.20.Uv, 68.37.-d

**Abstract**

Solution derived $La_2Zr_2O_7$ films have drawn much attention for potential applications as thermal barriers or low-cost buffer layers for coated conductor technology. Annealing and coating parameters strongly affect the microstructure of $La_2Zr_2O_7$, but different film processing methods can yield similar microstructural features such as nanovoids and nanometer-sized $La_2Zr_2O_7$ grains. Nanoporosity is a typical feature found in such films and the implications for the functionality of the films is investigated by a combination of scanning transmission electron microscopy, electron energy-loss spectroscopy and quantitative electron tomography. Chemical solution based $La_2Zr_2O_7$ films deposited on flexible Ni-5at.%W substrates with a {100}<001> biaxial texture were prepared for an in-depth characterization. A sponge-like structure composed of nanometer sized voids is revealed by high-angle annular dark-field scanning transmission electron microscopy in combination with electron tomography. A three-dimensional quantification of nanovoids in the $La_2Zr_2O_7$ film is obtained on a local scale. Mostly non-interconnected highly facetted nanovoids compromise more than one-fifth of the investigated sample volume. The diffusion barrier efficiency of a 170 nm thick $La_2Zr_2O_7$ film is investigated by STEM-EELS yielding a 1.8 ± 0.2 nm oxide layer beyond which no significant nickel diffusion can be detected and intermixing is observed. This is of particular significance for the functionality of $YBa_2Cu_3O_{7-\delta}$ coated conductor architectures based on solution derived $La_2Zr_2O_7$ films as diffusion barriers.

**Introduction**

Understanding the growth mechanisms and texturing of thin film coatings is of crucial importance for the fabrication of a variety of functional materials where thermal barrier coating, thermal resistance, coating toughness, diffusion barrier quality, crystallographic texture and electrical transport properties play an essential role. Oxide thin film coatings are interesting because of their numerous applications as catalysts, radiation resistant layers, electrolyte materials in solid oxide fuel cells, dielectric mediums for capacitors and as buffer layers for coated conductor architectures.[1] Chemical solution deposition[2] and sol-gel based processing[1] have emerged as promising routes for the fabrication of $YBa_2Cu_3O_{7-\delta}$ (YBCO) coated conductors.[2] These consist of a highly biaxially textured substrate upon which a buffer and superconducting layers can be deposited by a variety of film deposition techniques.

Coated conductors are of great interest for diverse applications in the energy and magnet technology; such as power cables, transformers and current limiters. [1] $La_2Zr_2O_7$ (LZO) thin films are currently of great interest for the fabrication of low-cost buffer layers because of their scalability, compatibility with the high critical current of YBCO, the small lattice mismatch of the a or b axis with that of YBCO (~0.5% and 1.8%, respectively), the relatively low formation temperature of ~900°C, the high stability of up to 1500 °C and the capability to grow biaxially textured on flexible nickel tungsten substrates[3], acting both as a Ni diffusion barrier layer and as a seeding for the upper layers. The buffer layers play an important role for coated conductor technology since they transfer the texture from the highly biaxially textured nickel tungsten substrate up to the YBCO superconducting layer.[4]

The microstructure of LZO thin films shows exciting properties, the layers are highly biaxially textured, but are nevertheless non-coherent[4]. Annealing and coating parameters strongly affect the microstructure, but different processing methods yield similar nanostructures. Nanoporosity with nanovoids of 10-20 nm in size and LZO grains (100-200 nm) are typical features found in such films [5]. Several film deposition techniques can be used for obtaining similar solution derived pyrochlore films.[1,6,7,8,9,10] Substrates can be flexible nickel tungsten or LaAlO$_3$ (LAO) single crystal substrates, but similar features have been observed and have been reported in the literature. [7] To avoid significant oxidation of the underlying substrate these oxide thin films are treated in a reducing gas (Ar-5%H$_2$). Such a deposition process yields a porous microstructure, so that chemically prepared LZO thin films can be considered as porous materials. Nanovoids are formed due to the combustion of organic material at the decomposition and this pyrolysis retains carbon in the material and seems to be unavoidable under the standard preparation conditions.[7],[4],[11]

The goal of the present investigation was to (i) demonstrate the efficiency of LZO buffer layers as diffusion barriers for Ni on the nanoscale in spite of the presence of nanoporosity and (ii) to perform a three dimensional nanoscale characterization by quantitative electron tomography. Reliable measurements of the diffusion barrier efficiency of chemically deposited LZO buffer layers and of the LZO nanoporosity were unavailable up till now. This is of great relevance for the quality control and functionality of chemically deposited buffer layers for coated conductor technology.[12] High-angle annular dark-field scanning transmission electron microscopy (HAADF-STEM) combined with electron energy-loss spectroscopy (EELS) has been performed to investigate the efficiency of the LZO buffer layer as a diffusion barrier at the nanoscale. Combining HAADF-STEM with electron tomography, the 3D morphology of the LZO thin film has been investigated. Using the reconstruction algorithm technique 'Discrete Algebraic Reconstruction Algorithm' (DART)[13],[14], the nanoporosity of the LZO buffer layer could be determined quantitatively.

**Experimental**

A biaxially-textured Ni-5at%W tape was dip coated using a (0.15 M) LZO solution, pyrolyzed and annealed at a temperature of 900°C. The pyrolysis was performed at 600°C with a heating rate of 10 K/min and under air. After pyrolysis the sample was brought to room temperature and annealed with a heating rate of 10 K/min. Annealing time was 60 min and cooling was done at 2-3 K/min. Details on sample preparation procedures are found in [31]. The LZO film thickness was 170 nm.

Samples for TEM investigations were prepared by conventional mechanical polishing and grinding followed by ion milling using a Res 100 Baltec ion milling machine operating at 4.5 kV and 3.5 mA for several hours. Plan-view samples were ion milled from the substrate side only at an angle of 12°. Cross-sectional samples were prepared by a grinding and polishing process and then ion milled from one side only at an angle of 12°. The final polishing stage was done at 6°. A micro-pillar was prepared for electron tomography using a a FEI Nova Nanolab 200 DualBeam SEM/FIB system, this allowed for a full tilt range avoiding missing wedge artefacts. [25]

HAADF-STEM and STEM-EELS was performed using a JEOL 3000 F microscope equipped with a Gatan GIF2000 1K spectrometer system operating at 300 kV with an energy dispersion of 0.5 eV/channel and an approximate energy resolution of 1.5 eV. Electron energy loss spectroscopy (EELS) scans in STEM mode were performed across the layer interface with a collection angle of 28.6 mrad and a convergence angle of 10.4 mrad. EELS spectra were analysed using Digital Micrograph and EELSMODEL software.[16] The sample was tilted into the [001] zone-axis of Ni to keep the interface parallel to the electron beam for the STEM-EELS measurements. Spatial drift correction is also applied.

For electron tomography the micro-pillar was mounted on a dedicated Fischione 2050 on-axis rotation tomography holder allowing 360° image acquisition. A series of 2D HAADF-STEM micrographs is recorded over a tilt range of 180° with 2° tilt increments using a JEOL 3000 F TEM operating at 300 kV. After alignment of the micrographs using a cross-correlation algorithm the 3D volume was reconstructed using SIRT [26] and DART [13],[14] algorithms. Visualization was done with Amira software. To gain 3D information on the density of the voids an additional segmentation step has to be taken.

**Results and discussion**

LZO films were prepared by chemical solution deposition on Ni-5at.%W substrates with an annealing temperature of 900°C. Sample preparation details are found in reference.[4] Due to the anisotropic surface energy of LZO, the preferred planes for nanovoids are the hexagonally close packed {111} surfaces in fcc, yielding octahedral structures clearly seen as rectangles in the plan-view image of figure 1(a) and in the TEM cross-section image shown in figure 1(b). This preferential orientation for pyrochlore LZO structures has been previously reported. [4],[15] Figure 1(a) is a HAADF-STEM image of the chemical solution derived LZO thin film in plan-view. The contrast in HAADF-STEM images is proportional to the atomic number (Z) and the thickness of the sample; therefore it is also known as Z-contrast imaging. Thus, the square-shaped dark areas, 5-20 nm in size, are nanovoids and the bright background is LZO. A preferential direction is observed and the edges of the rectangles are parallel to [100] and [010] as shown schematically in figure 1(c). The projection of the octahedrons is observed in plan-view. Figure 1(b) is the corresponding cross-sectional TEM bright-field image. Note the preferred nanovoid facet orientation of 45° with respect to the substrate interface. The octahedron is shown schematically in figure 1(d). Figure 2 is a high resolution TEM image of a region shown in figure 1(b).

By combining STEM with EELS, local chemical composition from a specific sample area can be obtained with a high spatial resolution. A series of 2D EEL spectra (10 × 40) were acquired across the LZO-Nickel tungsten interface area marked by the box in figure 3. Pixel size along the y direction is 0.2 nm and along the x direction 5 nm. Figure 4 shows the typical O-K (532 eV), La-$M_{4,5}$ (832 eV and 849 eV) and Ni-$L_{2,3}$ (855 eV and 872 eV) edges obtained across the $La_2Zr_2O_7$/Ni-5at.%W interface. The 3D EELS data were analyzed by Digital Micrograph (DM) and EELSMODEL.[16] Because the La-$M_4$ edge (849 eV) overlaps with the Ni-$L_3$ (855 eV) edge, only the La-M5 edge was taken into account to quantify the La concentration by DM. To overcome the overlap problem for Ni, an attempt was made to separate only the Ni-$L_2$ edge making use of EELSMODEL. To calculate the Ni concentration, EELSMODEL was applied to extract ony the Ni-$L_2$ edge while the background was estimated from the flat region between the Ni-$L_3$ and the Ni-$L_2$ edge for the intensity and the tail of the Ni-$L_2$ edge for the slope. The derived La, Ni, O elemental maps are given in figure 5(a-c). All of the maps show that the concentration changes gradually at the interface between LZO and the Ni substrate over a limited distance. Figure 5 (d) is the corresponding R(La-$M_5$)G(Ni-$L_2$)B(O-K) image, an oxide layer as small as 1.8 ± 0.2 nm could be determined. Shown in figure 6 are the profiles of the O, La and Ni elemental map together with the relative intensity ratios of La/O, Ni/O. They show a step at the interface which implies that an intermediate layer is formed in between. Intermixing occurs and La is present in this regime. The presence of W, as low as 5% in the nickel substrate, cannot be measured here. Its $M_{4,5}$ ionization edges (at 1872 eV and 1810 eV respectively) are beyond the energy-loss range for a spectrum with reasonable signal to noise ratio (SNR). However, from HAADF-STEM images the interface appears as a dark layer, which suggests a relative lack of heavy atoms such as tungsten at the interface. Figures 7 shows the changes of the Ni-$L_2$ edge across the interface, a 1.3 eV energy shift of the Ni-$L_2$ ionization edge was observed. This chemical shift implies that the nickel is oxidized to a higher valence in the interface layer.[17] The ionization edge shift can be calibrated from reference materials and is approximately 1 eV/valence for the Ni-$L_2$ and $L_3$ edge.[18] The 1.3 eV Ni-$L_2$ shift implies therefore a valence increase of more than 1.[19] From the shift of the Ni-L2 edge position and the change of its peak shape, it is certain that the Ni in the interlayer is oxidized. La is also present in this domain. The La/O atomic ratio in the intermixing layer is 2/10, which is 70% of that of the $La_2Zr_2O_7$ layer. This ratio is close to the experimental result (65% ± 5) of La/O as can be seen in figure 6. If the interlayer would be $La_2NiO_4$ the La/O atomic ratio in the intermixing layer would be 87.5% of the $La_2Zr_2O_7$ layer, which is much higher than what we observed (65%). Also, more than 1.3 eV energy shift of Ni $L_2$ edge is observed. This is more than the 0.2 eV shift from Ni to NiO reported by Potapov *et al* [34]. This suggests that probably the Ni is oxidized to a valence even higher than 2+. Both of these proofs lead to the suggestion that the interlayer is an intermixing of $LaNiO_3$ and $ZrO_2$. However, since this occurs in a reduced atmosphere, the probable phase is $La_2Ni_2O_5$, which is a reduced form of $LaNiO_3$ [35, 36]. Even though chemically deposited LZO thin films are porous materials, they act as efficient nickel ion diffusion barriers. An ultrathin oxide layer was formed at the nickel substrate interface.

The Ni oxidation takes place due to the influence of temperature and the presence of oxygen during sample preparation, where nickel ions diffuse into the LZO layer; however a LZO buffer layer of 170 nm in thickness is sufficient to prevent further nickel ion diffusion into the YBCO superconducting layer, thus acting as an efficient diffusion barrier. The influence of depositing a YBCO superconducting layer on top of the LZO buffer layer is still a partially open question. Nickel oxide layers of 10-30 nm in thickness have been previously observed after YBCO deposition[20],[21] so that the oxide layer clearly increases in size during YBCO deposition, but the mechanism of further oxidation is still unclear. Usually before depositing the final YBCO superconducting layer, a thin $CeO_2$ film is deposited on top of the LZO buffer layer to further protect the superconducting YBCO layer from oxygen diffusion [22], since this would affect the superconducting properties. In that case a $BaCeO_3$ thin intermediate layer has been reported to be formed under the YBCO layer. [23] Cloet *et al.*[20]

reported the presence of nickel oxide and some $NiWO_4$ areas at the interface by EDX in the TEM, however, measurements were done on a full YBCO coated conductor sample in which they attribute the presence of the nickel oxides solely to the growth of a YBCO layer on top. It fails to provide information on the nickel oxide layer thickness and diffusion barrier efficiency of the LZO film itself. Although pores are clearly visible in the TEM images provided, no information is provided on the nanoporosity, nor are nanovoids identified as such. Furthermore, EDX in conventional TEM does not have the same spatial resolution as EELS combined with HAADF-STEM has; in the latter sub nanometer and even atomic resolution is possible. [24]

Nanovoids are typical features present in chemical solution derived LZO films; these were first detected with transmission electron microscopy by Molina *et al.*[4] A HRTEM investigation of $La_2Zr_2O_7$ thin films was reported confirming these results;[11] however no information on $La_2Zr_2O_7$ buffer layer efficiency, nickel oxide layer formation or nanoporosity density was reported. The ultra-thin oxide layer can also be observed in the images reported in other contributions.[11,33] Zhao *et al.*[9] reported nanovoid densities in similar solution derived pyrochlore thin films measured from plan-view bright-field TEM images; however, quantification is difficult since information is taken from 2D images and therefore no reliable nanovoid densities could be reported. No information on intermediate layers was provided. (S)TEM is the most reliable way to detect the nanovoids since it probes the volume of the sample. Due to the advantages of Z-contrast imaging for differentiating between the material and vacuum, HAADF-STEM emerges as the ideal technique for imaging the nanovoids. Electron tomography in combination with HAADF-STEM allows for a true 3D characterization and since the volume of the sample is probed, the local nanovoid density can be determined with a high precision.

However, difficulties arise when performing electron tomography due to the influence of missing wedge artifacts and segmentation methods; this has been overcome by the use of an on-axis tomography holder and a novel electron tomography image reconstruction technique 'Discrete Algebraic Reconstruction Algorithm' DART.[13] To eliminate any missing wedge a micro-pillar specimen was prepared for electron tomography using a FEI Nova Nanolab 200 Dual Beam SEM/FIB system. A micro-pillar mounted on an on-axis rotation tomography holder allows image acquisition with a full tilt range and so minimizing artifacts.[25] Figure 8 (a) shows a Focused Ion Beam (FIB) prepared micro-pillar consisting of the Ni-W substrate, the LZO buffer layer and a Pt layer, intentionally deposited for protecting the LZO layer. A series of 2D HAADF-STEM images was recorded over a tilt range of 180° with 2° increments using a JEOL 3000 F (S)TEM operating at 300 kV. Figure 8 (b) shows a typical HAADF-STEM image of the LZO micro-pillar sample. The tilt series is acquired in high-angle annular dark-field STEM mode to avoid unwanted diffraction contrast. This yields information only from the nanovoids and the surrounding LZO material as a whole, however information about the internal structure of the LZO is lacking.

The 3D volume is reconstructed using different algorithms: SIRT [26] and DART.[13], [14] Figures 8(c) shows a xy-orthoslice through the 3D reconstruction and figure 8(d) an xz-orthoslice through the 3D reconstruction. A volume rendering of the LZO material combined with an isosurface of the voids is shown in figure 8(e). Results on chemically deposited $La_2Zr_2O_7$ thin films showed that nanovoids occupy > 20% of the $La_2Zr_2O_7$ sample volume investigated. Using manual segmentation the density determined was (22.6±2.0) %, by segmentation through thresholding a value of (23.8±2.0) % (both using SIRT) and by the more reliable DART algorithm, a density of 20.7±1.9% was determined. Details on the reconstruction technique and segmentation procedure applied to porous materials have been published in a recent contribution.[32] Determining the nanoporosity density is important for LZO buffer layer quality control and for LZO thin film thickness estimations.[12] These values provide reliable LZO volume densities and are of importance in an industrial environment since not all buffer layer samples can be prepared for cross-sectional TEM. Thus, a quick and reliable film thickness estimation technique is important as an alternative to thickness measurements by TEM. Ellipsometry is such a technique for which real porosity density values could be used.

Voids have also been reported in metal organic deposition (MOD) YBCO films.[27] Electron tomography could be implemented to study the three-dimensional distribution of artificial pinning centers in tailored YBCO coated conductors for magnet and energy applications. Reports found in the literature deal mostly with intrinsic pinning centers and quantification is based on conventional reconstructions techniques. In this case, the quantification of pinning relevant secondary phases is limited due to missing wedge artifacts and under sampling.[28-30]

**Conclusions**

The efficiency of an LZO buffer layer as a diffusion barrier has been investigated with electron energy-loss spectroscopy in a (scanning) transmission electron microscope. Despite the overlap of the Ni-$L_3$ edge and La-$M_4$ edge, the distribution of Ni, La and O with a high spatial resolution at the interface are successfully derived. A combination of nickel and lanthanum oxide, is found only in a 1.8 ± 0.2 nm thin interface layer, beyond which no significant further nickel diffusion could be determined. STEM-EELS results suggest that the interlayer is an intermixing of $La_2Ni_2O_5$ and $ZrO_2$. The presence of W, as low as 5% in the nickel substrate, cannot be measured here. The oxide layer formed is a product of this process, beyond this layer no further diffusion could be measured. LZO nanoporosity was determined in a quantitative manner using electron tomography. The use of DART provided the best result for measuring the porosity in LZO films. This method can also be applied to other porous materials. In the case of the LZO film investigated, the nanoporosity density determined by this method was $20.7 \pm 1.9\%$ of the total sample volume.

**Acknowledgements**

The authors acknowledge financial support from the European Union under the Framework 6 program under a contract for an Integrated Infrastructure Initiative (Reference No. 026019 ESTEEM). We thank K. Knoth, B. Holzapfel for LZO synthesis under the framework and funding of the Virtual Institute "Chemically deposited YBCO superconductors" of the Helmholtz Gemeinschaft. H. Tan acknowledges financial support by the Flanders Research Foundation FWO under project nr. G.0147.06N. L. Molina acknowledges financial support by funding from the European Research Council under the Seventh Framework Programme (FP7), ERC grant N°246791 – COUNTATOMS. Authors thank L. Rossou and S. Van den Broek for TEM specimen preparation. L. Molina acknowledges helpful discussions with O. Eibl.

**Figures**

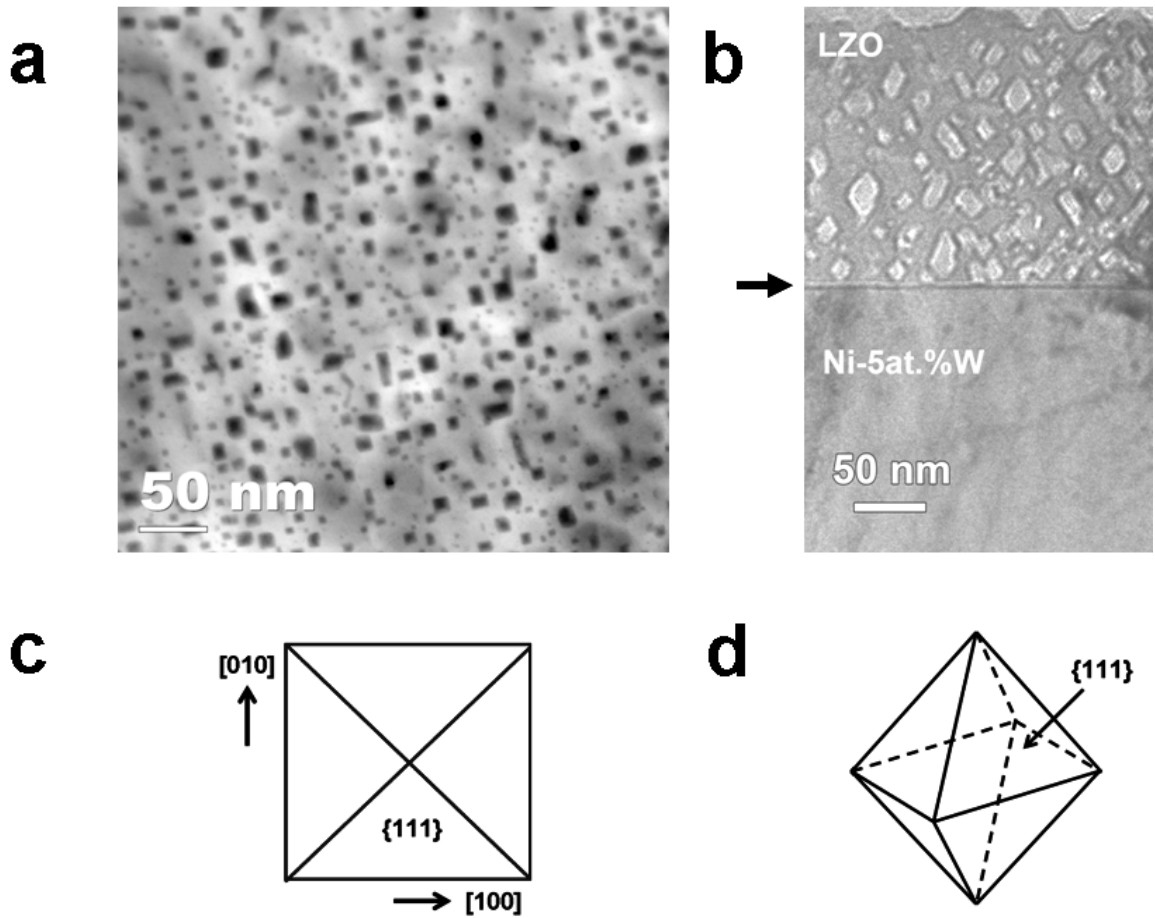

**Figure 1.** (a) HAADF-STEM image of a chemically deposited $La_2Zr_2O_7$ buffer layer in plan-view, seen from the <001> direction, a sponge-like structure is revealed. Rectangular shaped dark areas 5-20 nm in size are nanovoids, a preferential direction is observed and the edges of the rectangles are parallel to [100] and [010]. (b) Cross-sectional TEM bright-field image of a CSD deposited LZO thin film on a Ni-5at.%W substrate with a {100}<001> biaxial texture. The black arrow indicates the interface. Note the preferred nanovoid facet orientation of 45° with respect to the interface. Schematic drawings: (c) View from the <001> direction. The projection of the pyramids is observed in figure 1 (a). (d) The energetically preferred nanovoid structure has an octahedron shape in 3D.

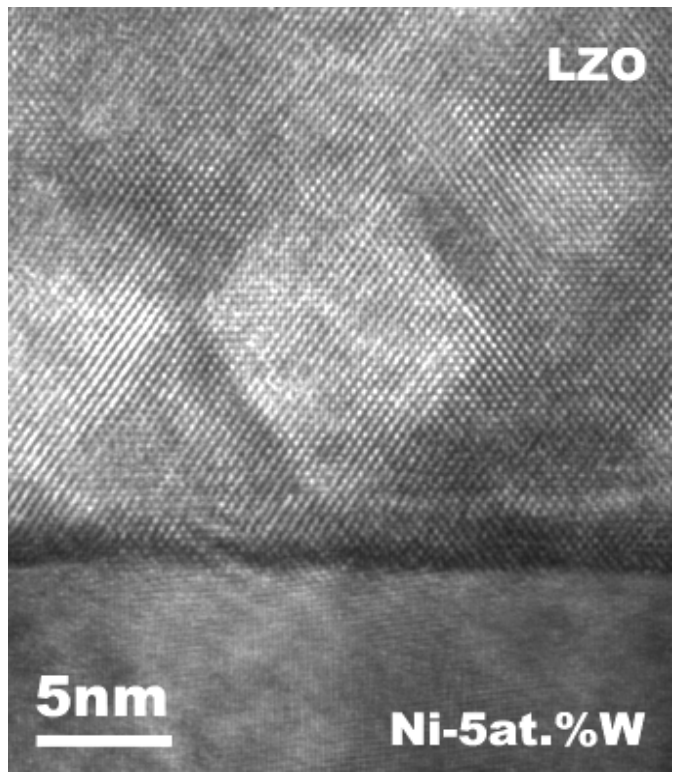

**Figure 2.** High resolution TEM image of the $La_2Zr_2O_7$/Ni-5at.%W interface. The facets of a nanovoid can be seen. Note the 2 nm layer at the interface.

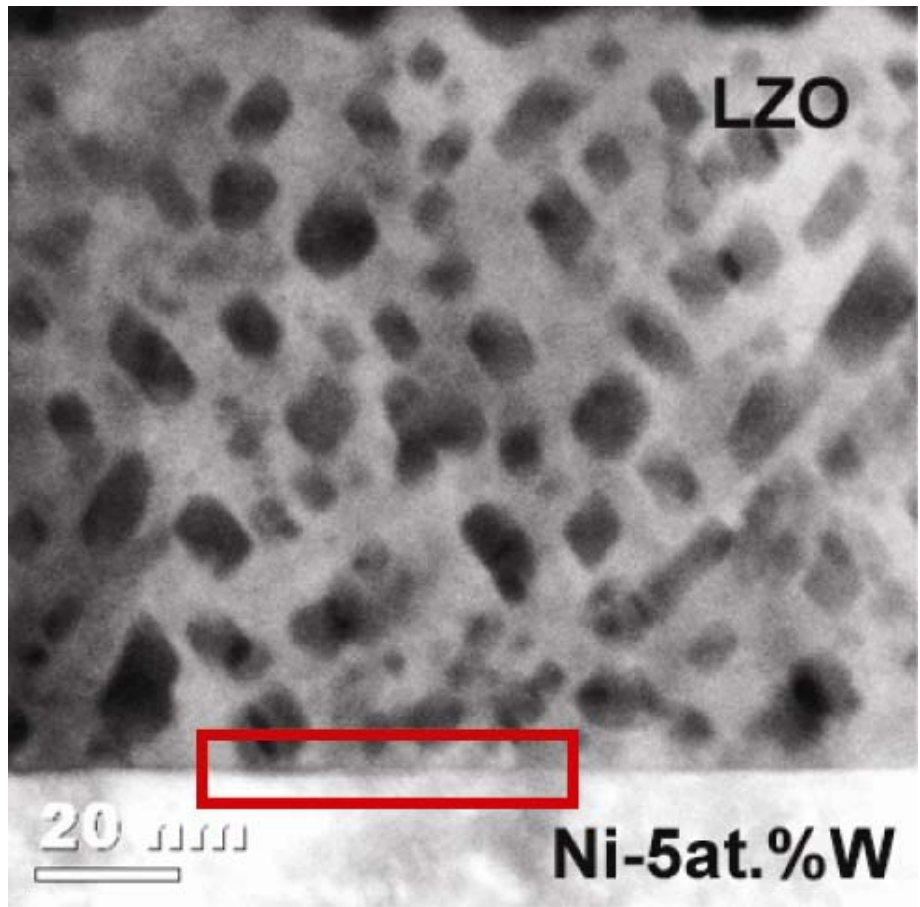

**Figure 3.** HAADF-STEM image of the $La_2Zr_2O_7$/Ni-5at.%W sample in cross-section. The red box indicates the area used for the STEM-EELS measurement. Nanovoids are observed in the $La_2Zr_2O_7$ layer.

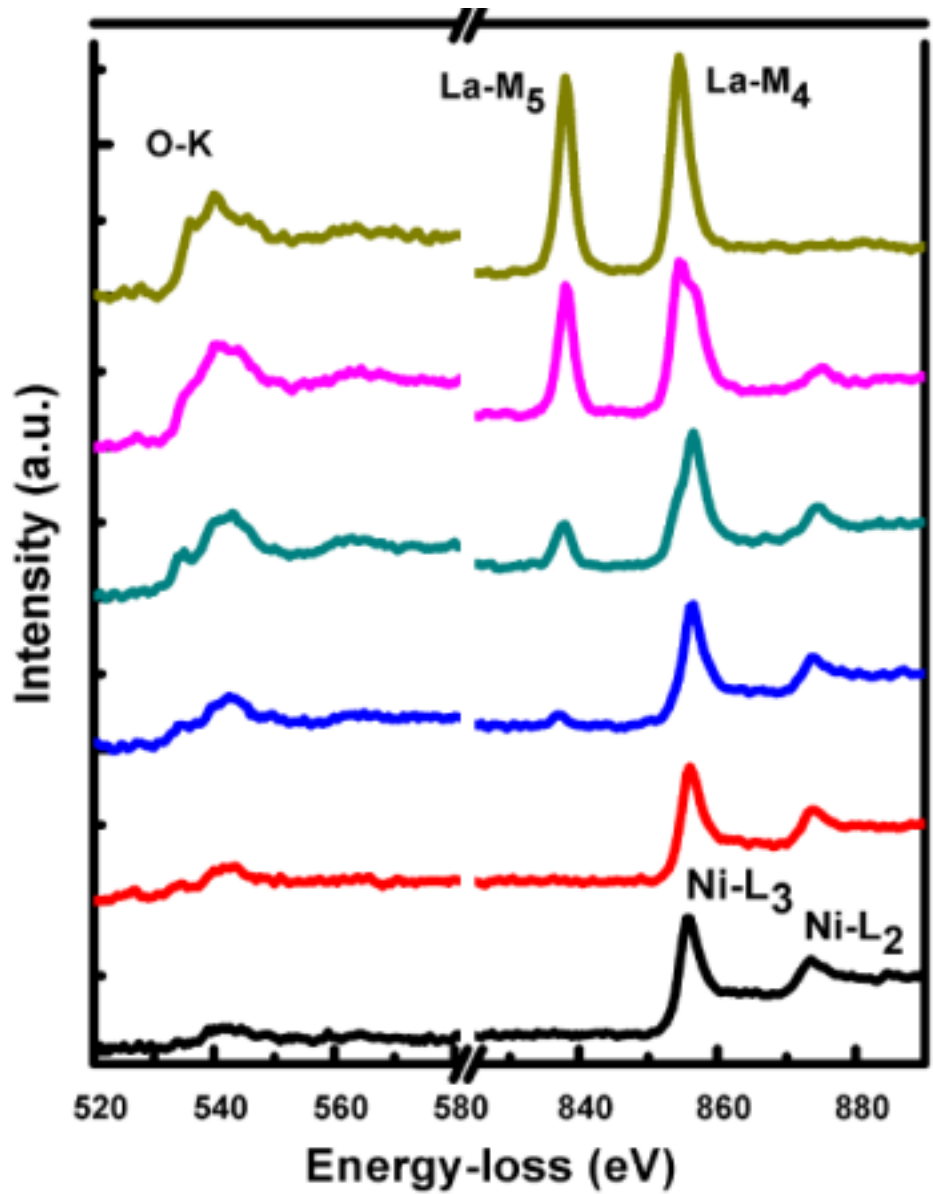

**Figure 4.** STEM-EELS spectra across the $La_2Zr_2O_7$/Ni-5at.%W interface. Spectra were obtained by averaging 10 spectra along the direction parallel to the interface. The material concentration changes from the nickel tungsten substrate to the nickel oxide layer at the interface, then to the $La_2Zr_2O_7$ buffer layer. Note that La-$M_4$ (849 eV) and Ni-$L_3$ (855) ionization edges overlap.

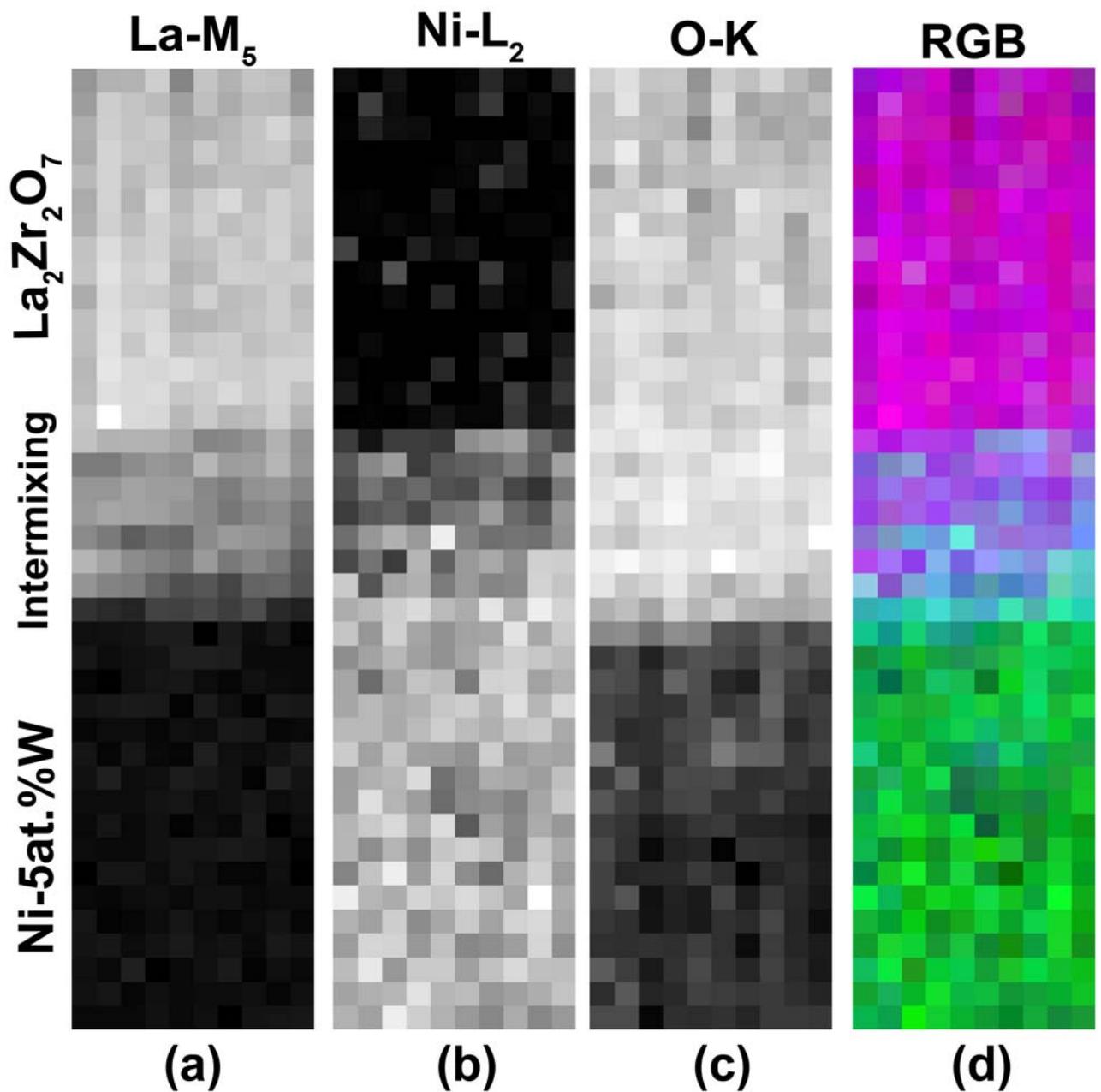

**Figure 5.** STEM-EELS 2D maps (10 x 40) pixels across the $La_2Zr_2O_7$/Ni-5at.%W interface (a) La-$M_5$ elemental map (b) Ni-$L_2$ elemental map (c) O-K elemental map and (d) the corresponding R(La-$M_5$)G(Ni-$L_2$)B(O-K) image. Pixel size along the y direction is 0.2 nm and along the x direction is 5 nm.

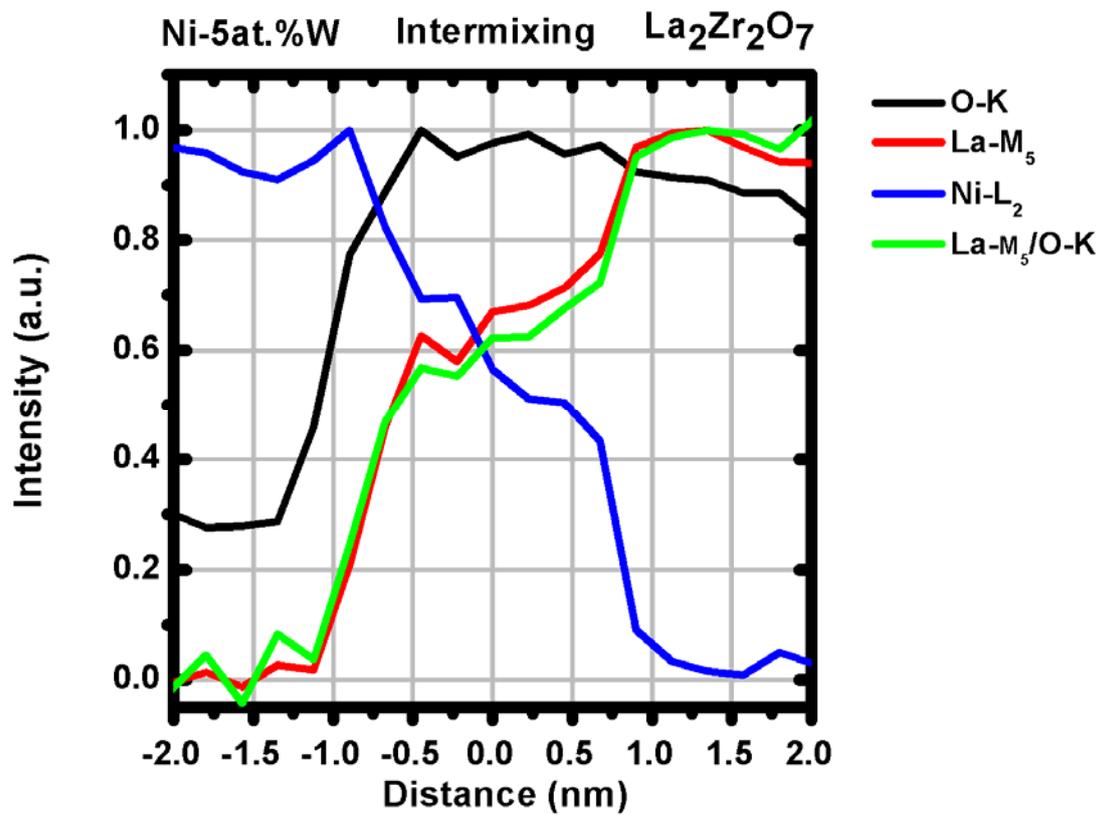

**Figure 6.** Integrated intensities for O-K, Ni-$L_2$ and La-$M_5$ ionization edges. The La-$M_5$/O-K ratio is shown.

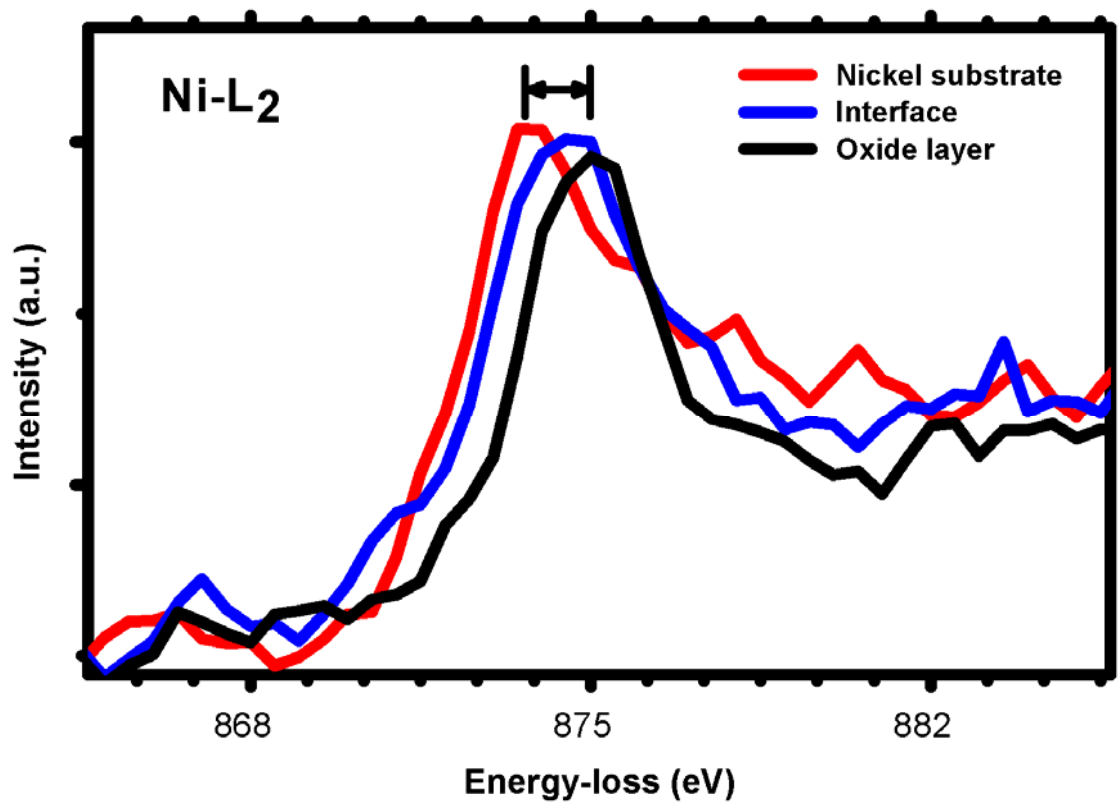

**Figure 7.** EELS Ni-L$_2$ spectra across the interface from the Ni-5at.%W substrate to the oxide layer. A 1.3 eV shift is observed.

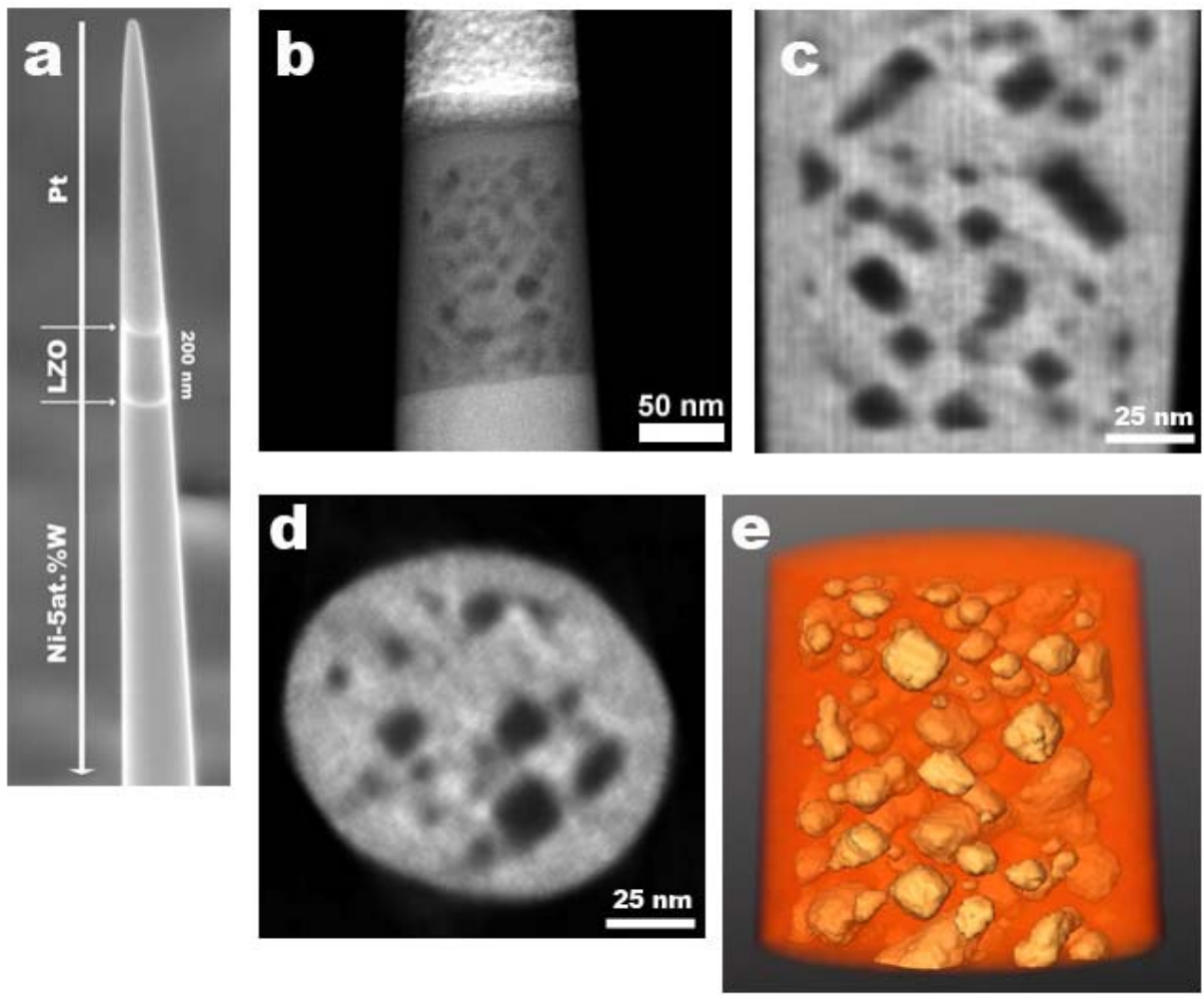

**Figure 8.** (a) Focused Ion Beam (FIB) prepared micro-pillar for on-axis rotation tomography (b) HAADF-STEM image of the LZO micro-pillar sample. (c) xy-orthoslice through the 3D reconstruction. (d) xz-orthoslice through the 3D reconstruction. (e) Volume rendering of the LZO material. Nanovoids are visualized.